\documentclass[aip,jap,amsfonts,amssymb,amsmath,groupedaddress,twocolumn]{revtex4}

\usepackage{graphicx}
\usepackage{amsmath}
\usepackage{amsfonts}
\usepackage{amssymb}
\usepackage{graphicx}
\usepackage{epstopdf}

\begin{document}

\title{Impact of dimensional crossover on phonon transport in van der Waals materials: a case study of graphite and graphene}

\author{Patrick Strongman}
\affiliation{Department of Physics and Atmospheric Science, Dalhousie University, Halifax, Nova Scotia, Canada, B3H 4R2}
\author{Jesse Maassen}
\email{jmaassen@dal.ca}
\affiliation{Department of Physics and Atmospheric Science, Dalhousie University, Halifax, Nova Scotia, Canada, B3H 4R2}

\begin{abstract}
Using first-principles modeling, we investigate how phonon transport evolves in layered/van der Waals materials when going from 3D to 2D, or vice versa, by gradually pulling apart the atomic layers in graphite to form graphene. Focus is placed on identifying the features impacting thermal conductivity that are likely shared with other layered materials. The thermal conductivity $\kappa$ of graphite is found to be lower than that of graphene mainly due to changes in the phonon dispersion driven by van der Waals coupling. Specifically, as the atomic layers are brought closer together, the acoustic flexural phonons in graphene form low-energy optical flexural phonons in graphite that possess lower in-plane velocities, density-of-states and phonon occupation, thus reducing $\kappa$. Similar dispersion changes, and impact on thermal conductivity, can be expected in other van der Waals materials when transitioning from 2D to 3D. Our findings also indicate that the selection rules in graphene, which reduce phonon-phonon scattering and contribute to its large $\kappa$, effectively hold as the atomic layers are brought together to form graphite. While the selection rules do not strictly apply to graphite, in practice similar scattering behavior is displayed due in part to the weak inter-layer coupling. This suggests that van der Waals materials, in bulk 3D form, may have lower phonon-phonon scattering rates than other non-layered bulk materials.
\end{abstract}

\maketitle

\section{Introduction}
Since the discovery of graphene, two-dimensional materials have been vigorously investigated due to their unusual and tailorable electronic, thermal, magnetic, optical and mechanical properties. The thermal transport characteristics of low-dimensional materials have been extensively explored \cite{Xu2016,Cocemasov2018,Zhang2020} for potential applications in heat management and thermoelectrics \cite{Song2018,Lindsay2018}, where very high or very low thermal conductivity $\kappa$ is desired, respectively. Often there are significant differences in heat transport when comparing a 2D material versus its 3D bulk counterpart. For example, the room temperature $\kappa$ of 3D graphite and 2D graphene are $\approx$\,2000 W/m-K and 2500-5000 W/m-K \cite{Nika2012,Wang2014,Nika2017}, respectively. Variations in $\kappa$, when comparing monolayer to multilayers and bulk, are common and have also been reported for black phosphorous \cite{Smith2017}, MoS$_2$ \cite{Gu2016}, GaN \cite{Jiang2017}, BN \cite{Sharma2020}, silicene and silicon \cite{Zhou2018,Fu2020}, among others.

With layered/van der Waals materials the difference between bulk and monolayer stems from weak inter-layer coupling. As a result the monolayer retains much of the bulk character, and vice versa, while displaying distinctive features. For instance, 2D materials possess flexural acoustic phonons with energy that scales quadratically $E(q)$\,$\propto$\,$q^2$ with phonon wavevector $q$ \cite{Carrete2016}, arising from a lack of atomic neighbors in the cross-plane direction, as opposed to the typical linear dependency $E(q)$\,$\propto$\,$q$. In the case of graphene, selection rules originating from its particular 2D symmetry prohibit anharmonic phonon-phonon scattering processes involving an odd number of flexural phonons \cite{Lindsay2010}. Such features influence $\kappa$ when a material transitions from 3D to 2D.

Previous studies have explored how the thermal characteristics of layered/van der Waals materials vary with thickness, by comparing the properties of monolayer, few-layer and bulk \cite{Smith2017,Gu2016,Jiang2017,Sharma2020,Zhou2018,Fu2020}. This approach, of changing the number of atomic layers, modifies the crystal symmetry and the phonon properties in a discrete and significant manner. Thus, it can be challenging to elucidate exactly how the thermal characteristics evolve from 3D to 2D, since there can be a significant leap between both limits.

This work focuses on understanding how phonon transport is modified during the transition from 3D to 2D (or vice versa) by taking a van der Waals material, in this case graphite, and gradually pulling the layers apart until they are effectively isolated graphene. While comparisons between mono-/few-layer graphene and graphite already exist \cite{Ghosh2010,Lindsay2011,Singh2011a,Singh2011b,Paulatto2013,Fugallo2014,Sun2019,Kim2019}, this approach provides a different perspective on how the phonon and scattering characteristics continously evolve from bulk to monolayer. We selected graphite/graphene as a case study, since it is one of the most common and extensively explored layered materials \cite{Aksamija2011,Zhong2011,Wei2011,Ong2011,Sadeghi2012,Qiu2012,Wei2012,Shen2013,Lindsay2014,Landon2014,Mei2014,Ni2014,Wei2014,Feng2015,Mu2015,Maurer2016,Majee2016,Karamitahari2016,Wang2017,Polanco2018,Zou2019}, and has shown interesting properties such as hydrodynamic phonon transport \cite{Lee2015,Cepellotti2015,Majee2018,Li2018,Li2019} and second-sound \cite{Huberman2019}. While this study focuses on one material system, effort is placed on identifying features that are common to a broader class of layered/van der Waals materials. This work is carried out using density functional theory (DFT) in conjunction with the phonon Boltzmann transport equation (BTE).

The paper is structured as follows. Section~\ref{sec:approach} introduces the theoretical and computational details, along with the selection rules that impact the phonon-phonon scattering rates. The results are presented in Sec.~\ref{sec:results}. In Sec.~\ref{sec:discussion} we summarize our findings and highlight the features likely shared with other layered materials. Finally, our conclusions are presented in Sec.~\ref{sec:conclusions}.

\section{Theoretical and computational approach} 
\label{sec:approach}
In this study we consider only the phonon contribution to thermal transport in graphite/graphene. The lattice thermal conductivity tensor elements, $\kappa^{\alpha\beta}$, are calculated using
\begin{align}
	\kappa^{\alpha\beta} = \frac{1}{\Omega N_q} \sum_{\lambda} v_{\lambda}^{\alpha} v_{\lambda}^{\beta} \tau_{\lambda} E_{\lambda}\frac{\partial f^0_{\lambda}}{\partial T}, \label{eq:kappa} 
\end{align}
derived from the linear phonon BTE \cite{Li2014}, where $\alpha$ and $\beta$ correspond to the Cartesian directions $x$, $y$ or $z$. Eq. (\ref{eq:kappa}) contains a sum over all phonon states $\lambda$\,=\,$(n,\mathbf{q})$, where $n$ and $\mathbf{q}$ are the phonon branch index and wavevector. $\Omega$ is the volume of the unit cell, $N_q$ is the total number of $\mathbf{q}$-points sampled in the Brillouin zone, $E_{\lambda}\equiv E(n,\mathbf{q})=\hbar \omega_{\lambda}$ is the phonon energy, $v_{\lambda}^{\alpha}=\partial \omega_{\lambda}/\partial q_{\alpha}$ is the phonon velocity, $f^0_{\lambda}$ is the Bose-Einstein distribution, and $\tau_{\lambda}$ is the scattering time. $\tau_{\lambda}$ captures the effect of 3-phonon scattering, including all normal and Umklapp processes, and must be calculated self-consistently (the theoretical details can be found in Ref.~\cite{Li2014}).

The thermal conductivity depends on two main components, the phonon dispersion and the scattering rates. The phonon eigen-energies and atomic eigen-displacements are obtained by diagonalizing the dynamical matrix, which depends on the second-order interatomic force constants (IFC). The scattering matrix elements for the 3-phonon transition rates are derived from the third-order IFCs. Thus, the second- and third-order IFCs are the fundamental inputs needed to carry out the thermal conductivity calculations. Next, we discuss how the IFCs are obtained from first-principles.

\subsection{Computational details}
\label{sec:compute}
Our computational approach is based on DFT in combination with the phonon BTE  \cite{McGaughey2019}. The DFT code Quantum Espresso (QE) \cite{QE1,QE2} was utilized to compute the second- and third-order IFCs, in conjunction with the ShengBTE \cite{Li2014} package that calculates the scattering rates, self-consistently solves the phonon BTE and returns $\kappa$. Density functional perturbation theory (DFPT) was used to calculate the second-order IFCs, and the finite difference method was used to extract the third-order IFCs.

For the QE calculations, the following simulation parameters were adopted. In the case of equilibrium and expanded graphite, we used a wavefunction energy cutoff of 60 Ry, a density energy cutoff of 400 Ry, a 15$\times$15$\times$3 $\mathbf{k}$-grid and a 6$\times$6$\times$3 $\mathbf{q}$-grid for the DFPT calculation. For graphene, a wavefunction energy cutoff of 70 Ry, a density energy cutoff of 2600 Ry, a 21$\times$21$\times$1 $\mathbf{k}$-grid and a 9$\times$9$\times$1 $\mathbf{q}$-grid for DFPT were adopted. In all cases, the effect of the core was treated with the projector augmented wave (PAW) method \cite{Blochl1994}, the exchange-correlation energy was included with the Perdew-Burke-Ernzerhof flavor of the generalized gradient approximation (PBE-GGA) \cite{Perdew1996}, and the van der Waals interaction was captured by the exchange-dipole model (XDM) \cite{Becke2007}. For the ShengBTE calculations, a $\mathbf{q}$-grid of 30$\times$30$\times$12 and 60$\times$60$\times$1 were adopted for graphite (and stretched graphite) and graphene, respectively, which were found to provide converged $\kappa$ values.

Atomic relaxations were carried out by minimizing the total energy and forces below their respective thresholds of $10^{-6}$ Ry and $10^{-5}$ Ry/Bohr. In the case of graphite, the structure was found to be unstable, giving negative phonon energies. Thus, the quasi-harmonic approximation was used to determine the optimal, and stable, structure at $T$\,=\,300~K. All scattering and thermal transport calculations were performed at $T$\,=\,300 K.

\subsection{Scattering selection rules}
\label{sec:selectrules}
While the main goal of this study is to identify generic features in the 3D to 2D transition of van der Waals materials, the case of graphene is unusual due its selection rules that forbid certain 3-phonon scattering processes \cite{Lindsay2010,Lindsay2011}. In this section, we briefly revisit the origin and consequences of the selection rules. We start by taking a Taylor expansion of the lattice potential energy $\Phi$ about its equilibrium value $\Phi_0$ for atomic displacements $\mathbf{u}_i$ away from the equilibrium positions $\mathbf{r}^0_i$:
\begin{align}
  		\Phi(\{\mathbf{r}^0\},\{\mathbf{u}\}) = &\,\,\Phi_0(\{\mathbf{r}^0\}) + \frac{1}{2}\sum_{ij,\alpha\beta} \Phi_{ij}^{\alpha\beta} u_i^{\alpha} u_j^{\beta} \nonumber \\ 
  		&+ \frac{1}{3!}\sum_{ijk,\alpha\beta\gamma} \Phi_{ijk}^{\alpha\beta\gamma} u_i^{\alpha} u_j^{\beta} u_k^{\gamma} + \cdots, \label{eq:potfull}
\end{align}
where the indexes $(i,j,k)$ each denote particular atoms, and $(\alpha,\beta,\gamma)$ each take on a Cartesian direction $x$, $y$ or $z$, $\Phi_{ij}^{\alpha\beta}\equiv [\partial^2\Phi/\partial u_i^{\alpha} \partial u_j^{\beta}]_0$ and $\Phi_{ijk}^{\alpha\beta\gamma}\equiv [\partial^3\Phi/\partial u_i^{\alpha} \partial u_j^{\beta} \partial u_k^{\gamma}]_0$ are the second- and third-order IFCs. The first-order term in Eq.~(\ref{eq:potfull}) is zero since it corresponds to evaluating the forces on the atoms at their equilibrium positions.

\begin{figure*}	
	\includegraphics[width=16cm]{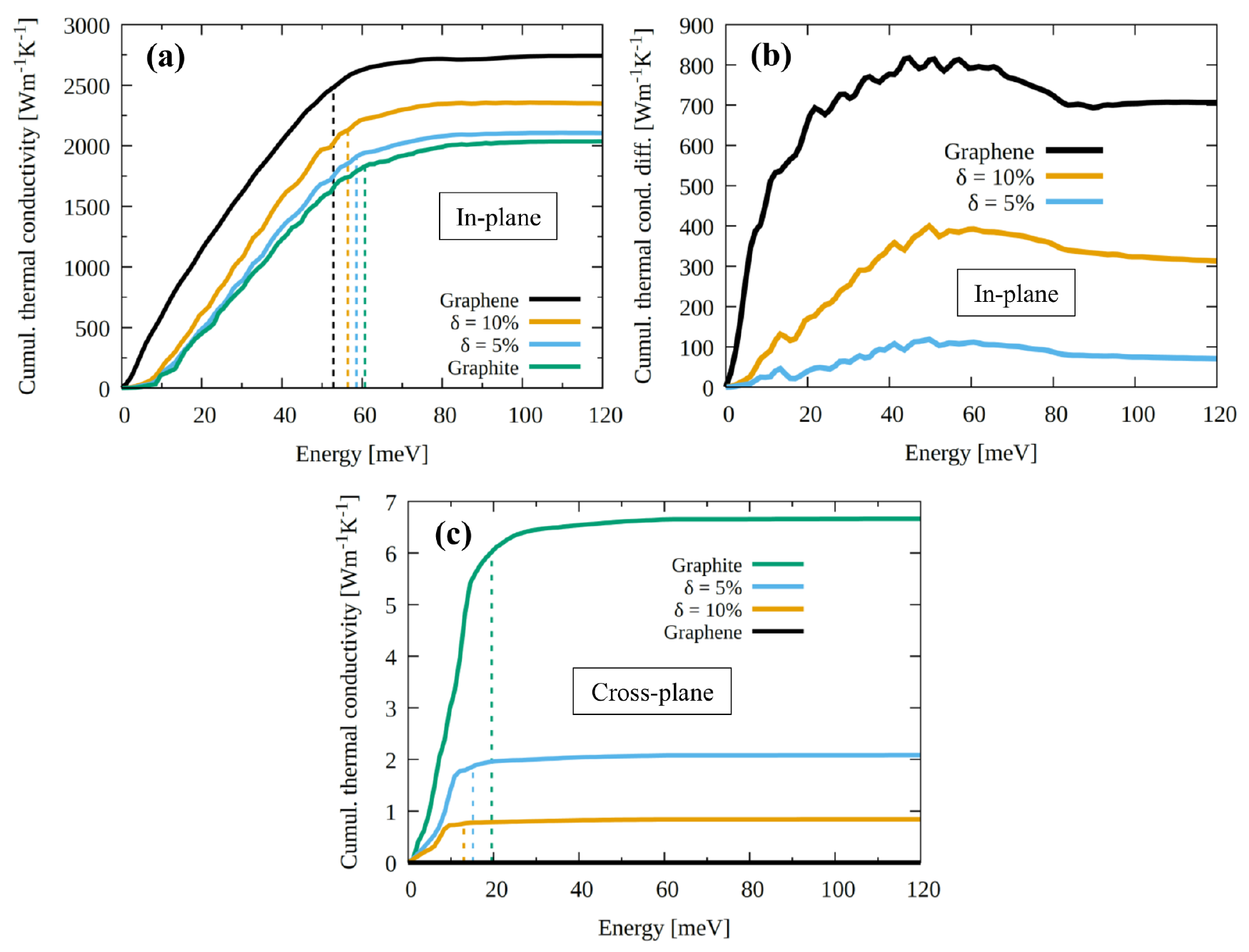}
	\caption{(a) Cumulative in-plane thermal conductivity versus phonon energy. (b) Difference in cumulative in-plane thermal conductivity relative to graphite. (c) Cumulative cross-plane thermal conductivity versus energy. Vertical dashed lines indicate when 90\% of the converged thermal conductivity is achieved. $T$\,=\,300 K.} \label{fig:cumulk}
\end{figure*}

The lattice potential given by Eq.~(\ref{eq:potfull}) must be invariant with respect to any symmetry operation. Graphene possesses a mirror symmetry about the plane of carbon atoms. Applying this symmetry to the second-order term in Eq.~(\ref{eq:potfull}) gives
\begin{align}
	\sum_{ij,\alpha\beta} \Phi_{ij}^{\alpha\beta} u_i^{\alpha} u_j^{\beta} = &\sum_{i'j',\alpha'\beta'} \Phi_{i'j'}^{\alpha'\beta'} u_{i'}^{\alpha'} u_{j'}^{\beta'} \nonumber \\
	= & \sum_{ij,\alpha\beta} (-1)^m \Phi_{ij}^{\alpha\beta} u_i^{\alpha} u_j^{\beta}, \label{eq:sumrule2}
\end{align} 
where all primed indexes correspond to those after the symmetry operation is applied, and $m$ refers to the number of occurrences of $z$ in $\alpha\beta$ (explained below). Since all the atoms are in the $x$-$y$ plane each atom maps to itself, $(i',j')=(i,j)$. All displacements along $x$ or $y$ are unchanged, $u_i^{x',y'}\rightarrow u_i^{x,y}$. However, any displacement along $z$ introduces a factor of $-1$, $u_i^{z'}\rightarrow -u_i^{z}$, thus explaining the factor $(-1)^m$ in Eq.~(\ref{eq:sumrule2}). The above condition requires that 
\begin{align}
	\Phi_{ij}^{\alpha\beta} = (-1)^m \Phi_{ij}^{\alpha\beta}, \label{eq:rule2}
\end{align}
which implies that $\Phi_{ij}^{\alpha\beta}$ must vanish for single $z$ component ($m=1$). It is straighforward to extend this analysis to higher-order order terms in Eq.~(\ref{eq:potfull}). For the third-order term, we find
\begin{align}
	\Phi_{ijk}^{\alpha\beta\gamma} = (-1)^m \Phi_{ijk}^{\alpha\beta\gamma}, \label{eq:rule3}
\end{align}
which states that $\Phi_{ijk}^{\alpha\beta\gamma}$ must vanish for an odd number of $z$ components. A consequence of Eq.~(\ref{eq:rule2}) is that all eigen-displacements of the atoms are either parallel or perpendicular to $z$ (those aligned along $z$ are called flexural). Finally, Eq.~(\ref{eq:rule3}) guarantees that any 3-phonon scattering process involving an odd number of flexural phonons is forbidden (see Appendix~\ref{app:selectrules} for details). The selection rules remove many of the possible scattering events, significantly lowering the total scattering rates and increasing the thermal conductivity of graphene.

In the case of graphite, the crystal structure is also mirror symmetric for an $x$-$y$ plane cutting through the atoms. The difference with graphite, however, is that when evaluating Eq.~(\ref{eq:sumrule2}) for atoms $i$ and $j$ in different layers the mirror symmetry does not map each atom onto itself. This reduces the number unique IFCs, but does not force them to vanish \cite{Lindsay2011}. Consequently, the eigen-displacements are not restricted to be parallel or perpendicular to $z$, and the selection rules for graphene do not apply to graphite. As will be shown later, while the selection rules are not strictly enforced in graphite, in practice the IFCs corresponding to scattering processes with an odd number of flexural phonons are nearly zero as a result of the weak inter-layer coupling.

\begin{figure*}	
	\includegraphics[width=17cm]{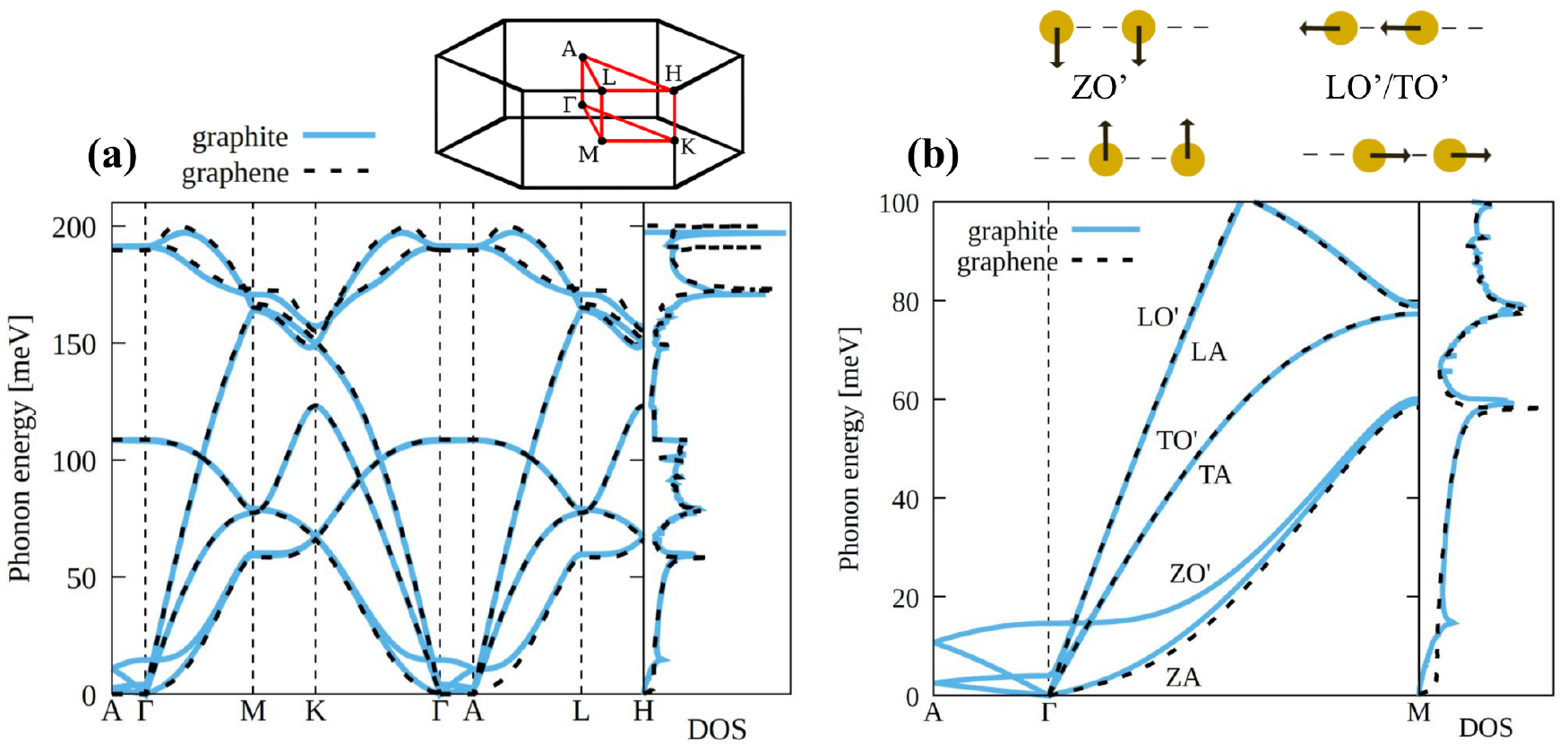}
	\caption{(a) Phonon energy dispersion, and density-of-states, of graphite and graphene. (b) Zoomed-in portion of dispersion and DOS in (a). Above panel (a): Brillouin zone of hexagonal graphite. Above panel (b): atomic displacements of ZO' and LO'/TO' modes in graphite.} \label{fig:dispersion}
\end{figure*}

\section{Results}
\label{sec:results}
As mentioned earlier, the focus of this study is to understand how both the phonon dispersion and scattering properties are modified in the transition from 2D to 3D, and how those changes impact thermal transport. In addition to investigating the limiting cases of graphene and graphite, we simulate ``stretched'' graphite in which the inter-layer distance is gradually increased. When pulling the atomic sheets apart, the in-plane lattice constant is kept the same as graphite, and the cross-plane lattice constant is varied according to $(1$+$\delta)\,c_0$, where $c_0$ is the equilibrium cross-plane lattice constant of graphite. In this work we present four cases: graphite, $\delta$\,=\,5\%, $\delta$\,=\,10\% and graphene. Stretching beyond 10\% makes the structure unstable, which prefers to form bilayer graphene. Smaller differences in $\delta$ lead to minor changes that can easily be extrapolated from the other cases. The optimized lattice constants for graphite are $a_0$\,=\,2.470\,\AA\space and $c_0$\,=\,6.872\,\AA, which are within 0.45\% and 2.4\% of the experimental values \cite{Baskin1955}, and give an equilibrium inter-layer spacing of $c_0/2$\,=\,3.436\,\AA. The lattice constants for graphene are $a_0$\,=\,2.464\,\AA\space and $c_0$\,=\,10\,$a_0$, where $c_0$ was chosen large enough so that the interaction between graphene layers in neighboring cells is negligible.

\subsection{Thermal conductivity}
\label{sec:kappa}
Fig.~\ref{fig:cumulk}a shows the in-plane cumulative thermal conductivity, $\kappa^{\rm cumul}(E)$\,=\,$\int_0^E \kappa(E')\,dE'$, for graphite, stretched graphite and graphene. Thermal conductivity is normalized to the equilibrium inter-layer spacing. $\kappa^{\rm cumul}(E)$ displays the contribution of phonons at various energies. The converged values are $\kappa_{\rm graphite}$\,=\,2030~W/m-K, $\kappa_{5\%}$\,=\,2100~W/m-K, $\kappa_{10\%}$\,=\,2340~W/m-K, and $\kappa_{\rm graphene}$\,=\,2730~W/m-K, which show a monotonic increase as the layers are pulled apart. The thermal conductivity of graphite agrees very well with previous calculations \cite{Fugallo2014,Sun2019} and measurements \cite{Ho1974}. The thermal conductivity of graphene falls within the range of calculated \cite{Lindsay2010,Singh2011a,Paulatto2013,Lindsay2014,Fugallo2014,Wang2014} and measured \cite{Ghosh2010,Nika2012,Nika2017} values, which generally spans from 2500~W/m-K to 5000~W/m-K. The experimental $\kappa$ for suspended graphene has a fairly wide range, since it is sensitive to sample quality and environment. The theoretical values of $\kappa$ can vary by a factor of two depending on the particular choice in DFT method \cite{Qin2018}. Also, recent additional consideration of 4-phonon scattering has shown to sizably reduce $\kappa$ in graphene \cite{Feng2018,Gu2019}.



\begin{figure*}	
	\includegraphics[width=15cm]{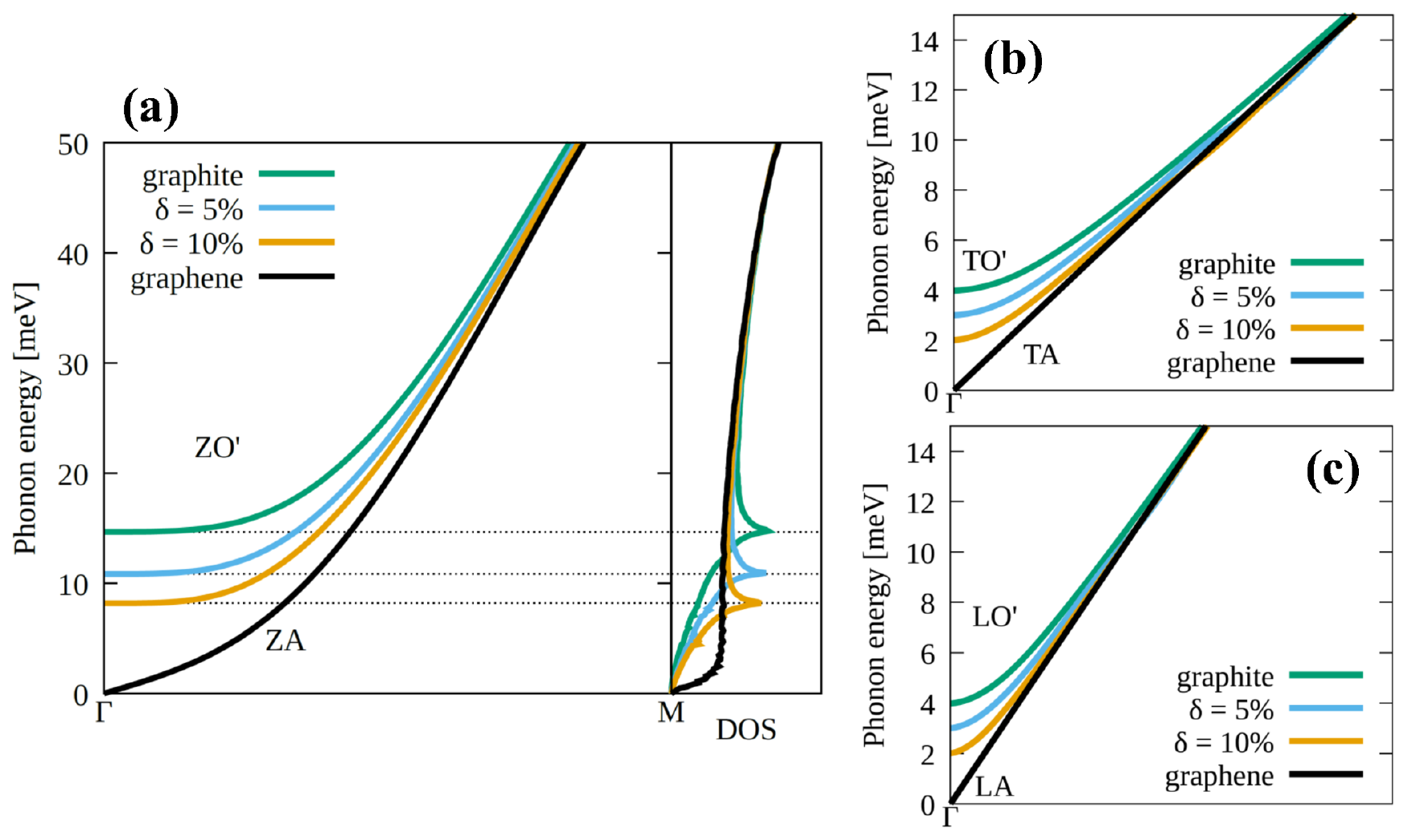}
	\caption{Phonon energy dispersion of ZO' (a), TO' (b) and LO' (c) branches for graphite and stretched graphite. These optical modes (ZO', TO', LO') in graphite become the corresponding acoustic modes (ZA, TA, LA) in graphene, as the inter-layer coupling vanishes. The graphite acoustic branches have been removed for clarity.} \label{fig:dispzoom}
\end{figure*}

From Fig.~\ref{fig:cumulk}, the largest contribution to $\kappa$ occurs in the range 10-50~meV in all cases, however with graphene there is a marked increase below $\approx$20~meV. To highlight this distinction, Fig.~\ref{fig:cumulk}b presents the difference in $\kappa^{\rm cumul}(E)$ relative to that of graphite. It is clear that graphene conducts much more than graphite in the 0-20~meV range, which accounts for most of the 700~W/m-K difference in converged $\kappa$. It was previously reported that the acoustic flexural phonons in graphene are responsible for roughly 75\% of $\kappa$ \cite{Lindsay2010}, which explains the rapid increase in $\kappa^{\rm cumul}(E)$ below 20~meV. Stretched graphite shows a larger contribution at low-energy, as the inter-layer distance increases, which appears to bridge both limits.

The cross-plane $\kappa^{\rm cumul}(E)$ is presented in Fig.~\ref{fig:cumulk}c. One noticeable feature is that the cross-plane $\kappa^{\rm cumul}(E)$ reaches 90\% of its converged value near 15-20~meV (indicated with vertical lines), as opposed to around 50-60~meV along in-plane. We also observe a much more dramatic change in cross-plane thermal conductivity as graphite is expanded; 6.7~W/m-K for graphite, 2.1~W/m-K for 5\% stretched and 0.9~W/m-K for 10\% stretched. The cross-plane $\kappa$ of graphite, 6.7~W/m-K, is consistent with other theoretical \cite{Fugallo2014,Sun2019} and experimental \cite{Ho1974} values.


%
%

Having presented how thermal transport varies from graphite to graphene, next we focus on understanding what causes these differences. Thermal conductivity depends generally on two factors, a material's {\it phonon dispersion} and {\it scattering properties} -- how both change in the crossover from 3D to 2D, and the impact this has on $\kappa$, is elucidated. Note that the results focus on energies below 100~meV, where the contributions to thermal conductivity are most significant.

\subsection{Properties related to phonon dispersion}
\label{sec:dispersion}
The phonon dispersion and density-of-states (DOS), for graphite and graphene, are shown in Fig.~\ref{fig:dispersion}a. These results agree well with previous calculations \cite{Paulatto2013,Lindsay2014,Sun2019} and measurements \cite{Maultzsch2004,Mohr2007}. Upon inspection it is apparent that the phonon energies are nearly identical for graphite and graphene, except at low energies where the weak inter-layer coupling plays a role. Fig.~\ref{fig:dispersion}b shows a closer view of the low-energy states. We note a few distinct features. With graphene, we find the acoustic flexural (ZA) phonons have a quadratic dispersion, which for a 2D material gives a constant DOS. The calculated DOS is roughly constant below 50~meV, but drops abruptly to zero near $E=0$. The transverse (TA) and longitudinal (LA) acoustic branches have a linear dispersion, resulting in DOS\,$\propto$\,$E$. With graphite, the ZA branch appears quadratic but becomes linear below a few meV. For a 3D material the DOS scales as $E^2$ and $E^{1/2}$ for linear and quadratic dispersions, respectively -- the former seeming closer to the calculated DOS.

\begin{figure*}	
	\includegraphics[width=12cm]{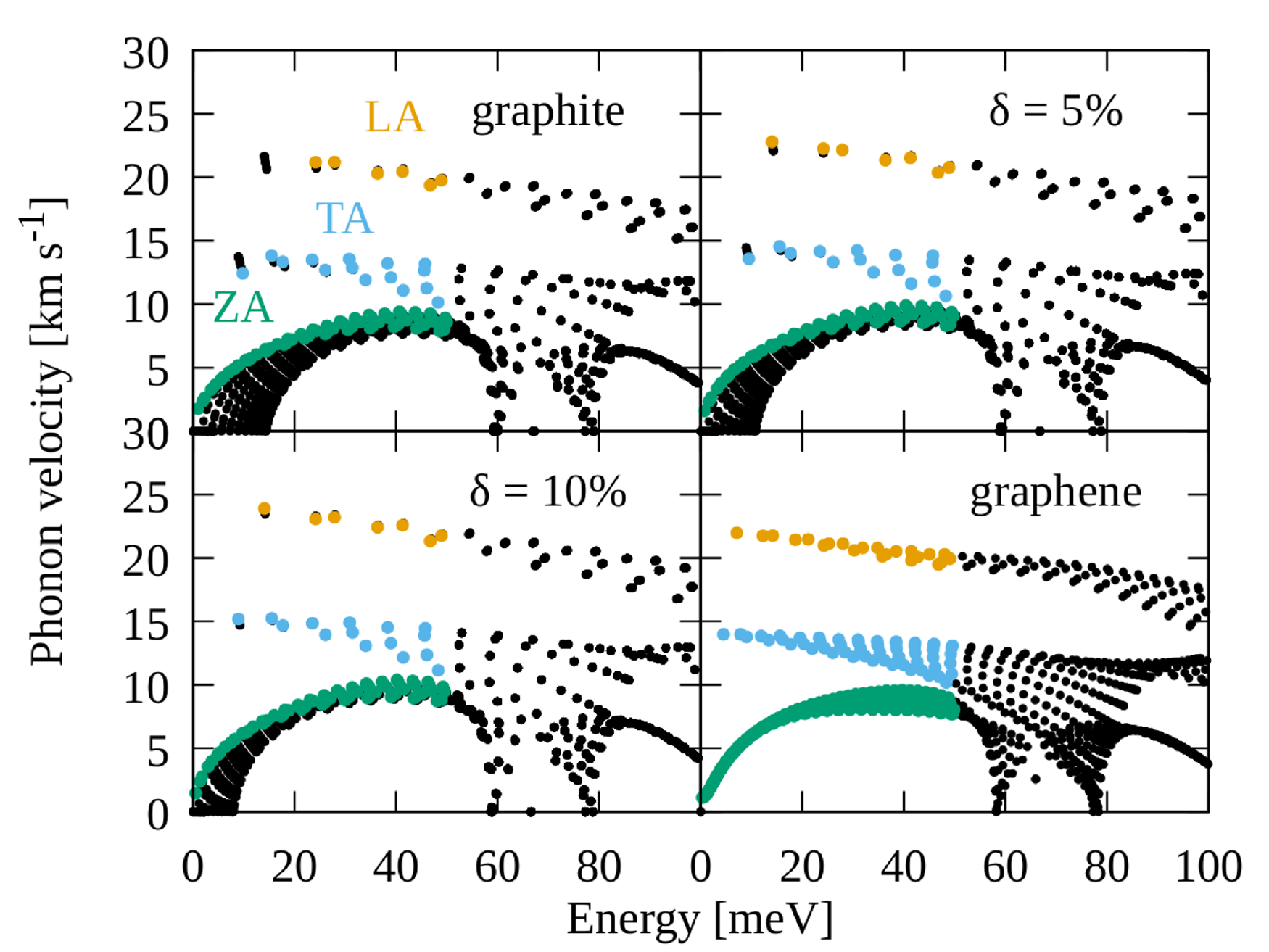}
	\caption{In-plane velocity magnitude versus phonon energy, for graphite, stretched graphite and graphene.} \label{fig:invelo}
\end{figure*}

Since the primitive cell of graphite contains twice the number of carbon atoms than graphene, the number of phonon branches is doubled (12 for graphite; 6 for graphene). The three acoustic branches (ZA, TA, LA) are nearly identical in both materials, as observed in Fig.~\ref{fig:dispersion}b. In graphite, there are three low-lying optical phonons (ZO', TO', LO'), not present in graphene, that result from the weak inter-layer coupling -- they arise from the interaction between the acoustic graphene states. The ZO' mode splits from the ZA below $\approx$30~meV, and remains quite flat while approaching $\mathbf{q}=0$, giving a peak in the DOS near 15~meV. As depicted above Fig.~\ref{fig:dispersion}b, ZO' results from two atomic layers oscillating towards and away from each other (derived from two out of phase ZA graphene modes). The TO' and LO' modes also closely follow the TA and TO states, respectively, but split away at low energy and cross $\Gamma$ at 4~meV. These low-lying optical phonons correspond to ``sliding'' modes in which both layers oscillate in-plane along opposite directions.

Fig.~\ref{fig:dispzoom} presents how the ZO', TO' and LO' branches in graphite evolve as the layers are streched apart to become graphene. Since these modes result from inter-layer van der Waals couping, their energy gradually decreases as the inter-layer spacing is increased. For a 10\% increase in inter-layer distance, the optical phonon energies drop by roughly half. The peak in DOS arising from the ZO' mode follows its value at $\Gamma$. As the layers continue to be pulled apart, the peak in DOS will approach $E\rightarrow0$ and flatten, resulting in the constant DOS of graphene. Moreover, as the branches drop in energy to join the acoustic graphene states their  curvature will increase, as will their velocities.

From the dispersion, the phonon velocity is obtained using $\mathbf{v}_{\lambda}$\,=\,$\partial \omega_{\lambda}/\partial \mathbf{q}$. Fig.~\ref{fig:invelo} shows the in-plane velocity magnitude for graphite, stretched graphite and graphene. The linear dispersion LA and TA branches approach constant velocities as $E\rightarrow0$, while the velocities of the ZA branch drop to zero. A prominent feature is how the ZA velocities spread out at low energy ($<$20~meV) as the layers are brought closer together. This occurs due to the inter-layer coupling that transforms graphene's 2D dispersion, which is independent of $q_z$, into graphite's 3D dispersion that depends on $q_z$ (see Fig.~\ref{fig:dispersion}). The ZA states with $q_z$\,$\neq$\,0 are shifted up in energy, hence their velocities are roughly the same as those for $q_z$\,=\,0 (shown in green) but shifted to the right in Fig.~\ref{fig:invelo}. The opposite is true for the ZO' states -- a $q_z$\,$\neq$\,0 component lowers the energies. As a result, the observed spread in velocities from the flexural states ranges in energy from zero to $E_{\rm ZO'}(\mathbf{q}=0)$. This feature gradually increases as the layers are brought together, since it originates from the inter-layer coupling, and lowers the average in-plane velocity at a given energy.

\begin{figure*}	
	\includegraphics[width=12cm]{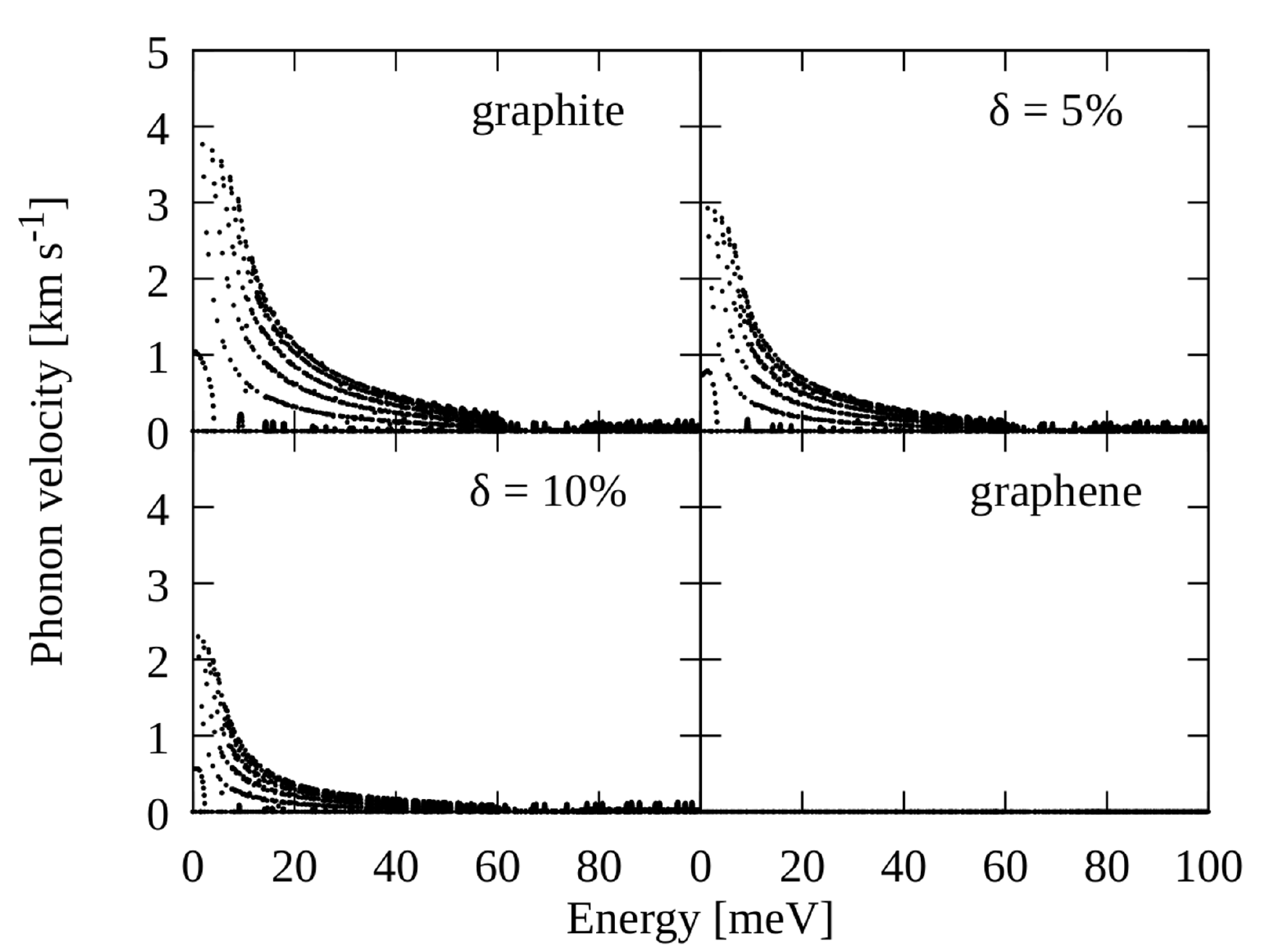}
	\caption{Cross-plane velocity magnitude versus phonon energy, for graphite, stretched graphite and graphene.} \label{fig:outvelo}
\end{figure*}

Fig.~\ref{fig:outvelo} presents the cross-plane velocity magnitude, which are significantly lower than along in-plane. The cross-plane velocities decrease rapidly with energy, unlike the in-plane velocities, since the inter-layer coupling only impacts the low-energy portion of the dispersion. As the atomic sheets are expanded, the cross-plane velocities drop significantly and tends to zero as the van der Waals interaction vanishes. For 10\% increase in inter-layer spacing, the maximum velocity is lowered by almost half.

Having analyzed the properties related to phonon dispersion, and how they change from graphite to graphene, we briefly discuss how this influences thermal conductivity. i) The ZO' phonons in graphite (and stretched graphite) have larger energies than their equivalent ZA phonons in graphene. From Eq.~(\ref{eq:kappa}) we see that $\kappa$ is proportional to the factor $E\,\partial f^0/\partial T$, a monotonically decreasing function, which is approximately equal to $k_B[1-E^2/(4k_BT)^2]$ in the limit $E\ll k_BT$. Thus, the low-lying ZO' phonons contribute less to $\kappa$ than the ZA phonons, from an occupation perspective. ii) The inter-layer coupling reduces the average in-plane velocity of the flexural phonons in graphite, thus lowering $\kappa$ relative to graphene. iii) The constant DOS of graphene leads to far greater number of phonons below $\approx$\,10~meV, compared to graphite, that can contribute to heat flow. iiii) The cross-plane velocities decrease rapidly with energy, unlike in-plane, which explains why the cross-plane $\kappa^{\rm cumul}(E)$ converges near $\approx$\,20~meV. In contrast, the in-plane $\kappa^{\rm cumul}(E)$ saturates near $\approx$\,60~meV due to the exponential decrease in phonon occupation -- the factor $E\,\partial f^0/\partial T$ is approximately equal to $k_B(E/k_BT)^2\exp{(-E/k_BT)}$ in the limit $E\gg k_BT$. Previous studies have pointed out how the low-energy ZO' modes in multi-layer graphene lowers the velocities and reduces $\kappa$, compared to graphene \cite{Ghosh2010,Singh2011b}.

Next, we investigate how the 3-phonon scattering properties vary from graphite to graphene, and the impact on thermal transport.

\subsection{Properties related to phonon-phonon scattering}
\label{sec:scat}
\begin{figure*}	
	\includegraphics[width=13cm]{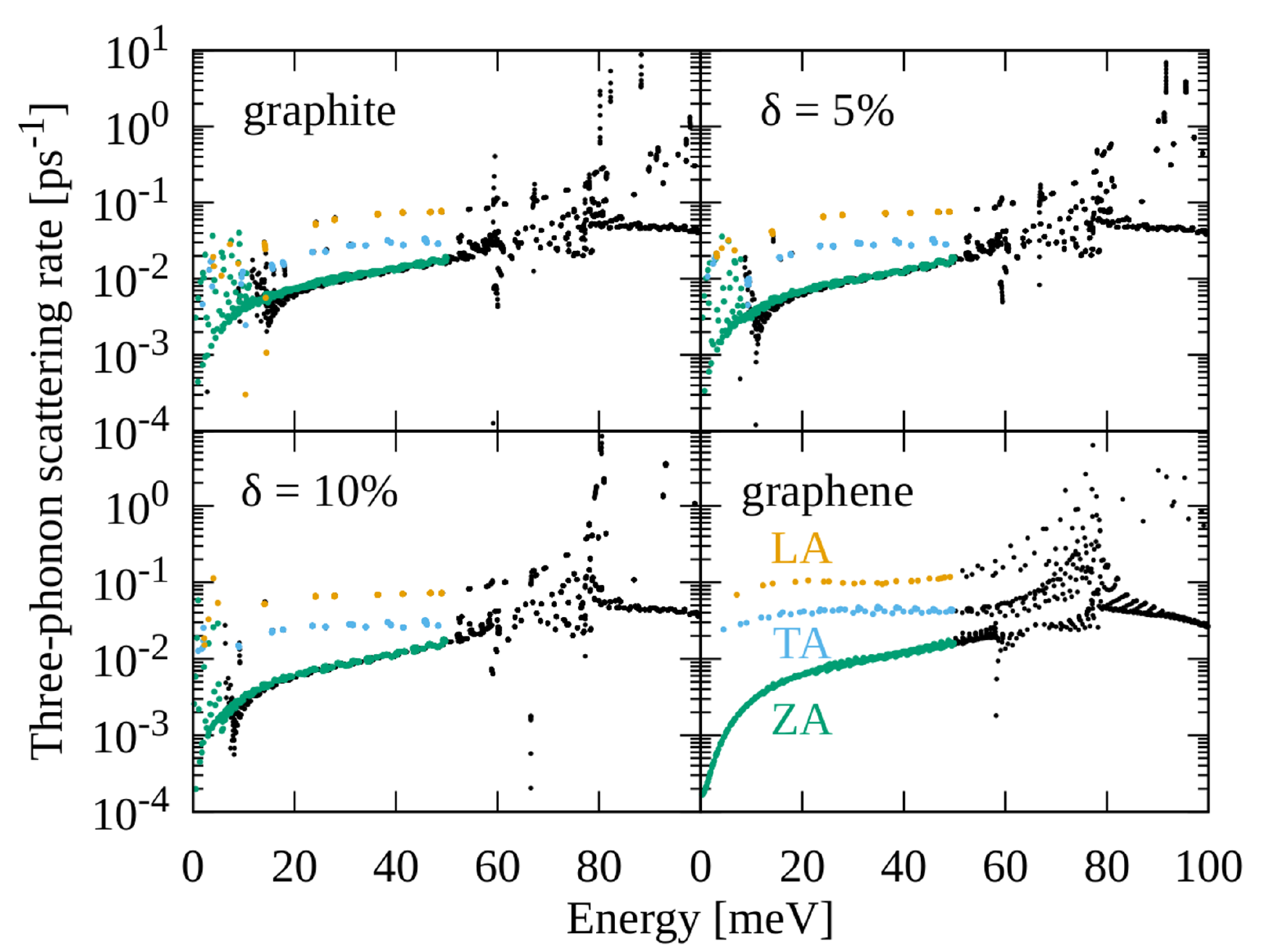}
	\caption{Converged three-phonon scattering rates, $\tau^{-1}_{\lambda}$, versus phonon energy, for graphite, stretched graphite and graphene. $T$\,=\,300 K.} \label{fig:scatrate}
\end{figure*}

The 3-phonon scattering rates for graphite, stretched graphite and graphene are presented in Fig.~\ref{fig:scatrate}. In all cases the scattering rates, $\tau^{-1}_{\lambda}$, for the LA and TA modes are largest below 60~meV (the energy range relevant to transport). With graphene, $\tau^{-1}_{\lambda}$ is noticeably lower for the ZA phonons, especially $<$\,10~meV -- a result of the scattering selection rules \cite{Lindsay2010}. As the graphene layers are brought closer together, we observe a spread in $\tau^{-1}_{\lambda}$ at low energies. This arises from the 3D phonon states with $q_z$\,$\neq$\,0 discussed earlier. In addition, there is a dip in $\tau^{-1}_{\lambda}$ at the energy corresponding to $E_{\rm ZO'}(\mathbf{q}=0)$. Aside from these differences, the scattering rates are fairly similar in all cases. For example, both graphite and graphene show scattering rates for the flexural modes reaching as low as $\approx$\,10$^{-3}$\,ps$^{-1}$. With 10\% cross-plane compression in graphite, the scattering rates rise only slightly at low energy \cite{Sun2019}, which is consistent with our findings. The mean-free-paths are also fairly similar in graphite and graphene (see Fig.~\ref{fig:mfp} in Appendix). This, however, is unexpected since the scattering rates for ZA phonons in graphene are anticipated to be much lower than graphite, since the selection rules do not apply to graphite.

\begin{figure}	
	\includegraphics[width=8.7cm]{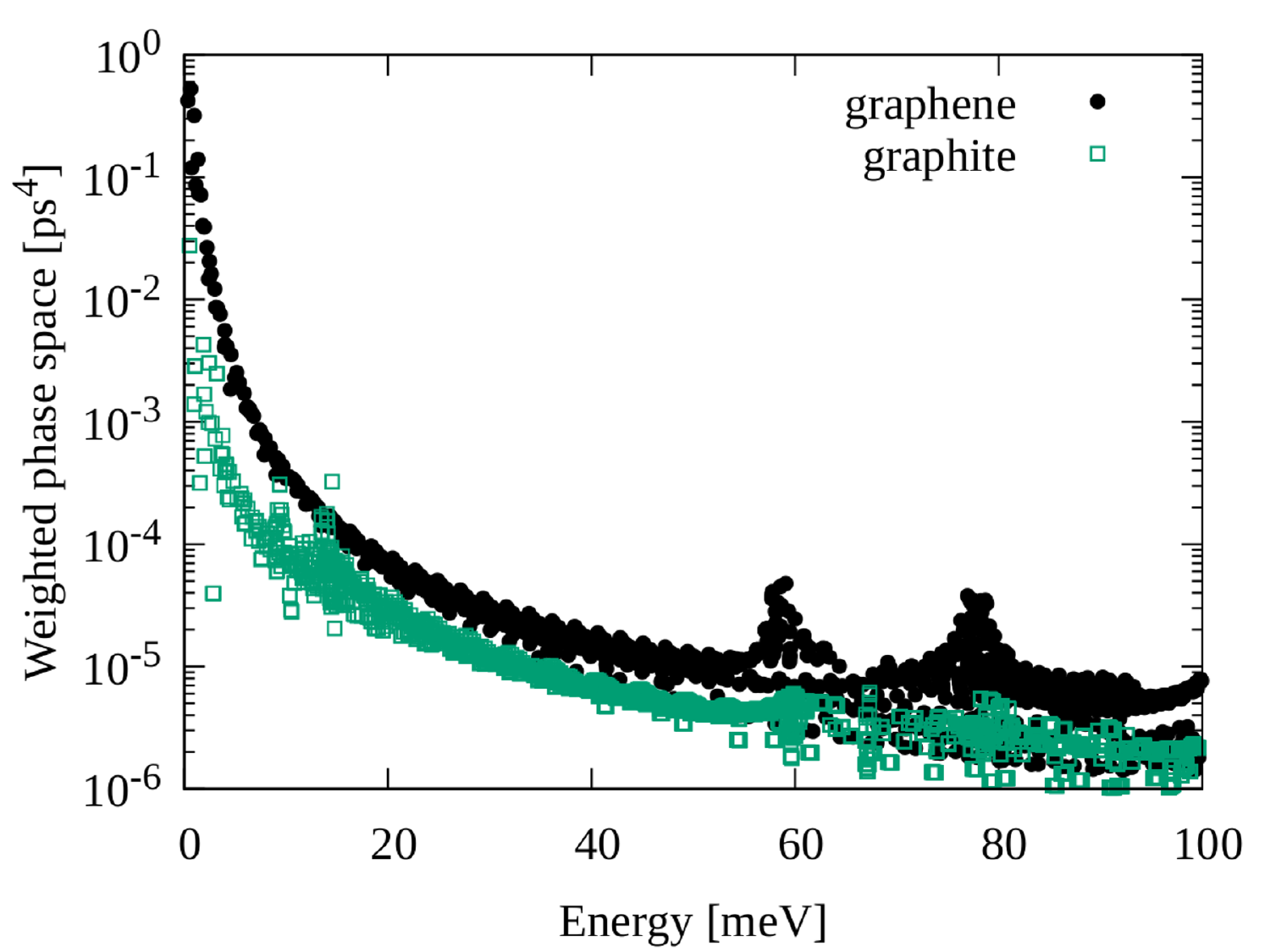}
	\caption{Weighted phase space, $W_{\lambda}$, versus phonon energy for graphite and graphene. $T$\,=\,300 K.} \label{fig:wps}
\end{figure}

To better understand the behavior of the scattering properties, we can write the scattering rates as the product of two physically-intuitive terms:
\begin{align}
       \frac{1}{\tau^0_{\lambda}}=\frac{h}{8}\,\langle |V_{\lambda}|^2 \rangle W_{\lambda}, \label{eq:scat_VW}
\end{align}
where $1/\tau^0_{\lambda}$ are the unconverged 3-phonon scattering rates (corresponding to the initial guess for the self-consistently solved scattering rates $1/\tau_{\lambda}$ and solution to the relaxation time approximation), $\langle |V_{\lambda}|^2 \rangle$ is the average squared scattering potential and $W_{\lambda}$ is the weighted phase space \cite{Li2015} (details on the definitions of these quantities are found in Appendix~\ref{app:wps_avpot}). $W_{\lambda}$ is a measure of how many scattering processes are possible, given the constraints of energy and crystal momentum conservation, and depends only on the phonon energies and occupations. $\langle |V_{\lambda}|^2 \rangle$ is a weighted average of the 3-phonon scattering matrix elements over all the processes counted in $W_{\lambda}$, and thus captures the effect of the third-order IFCs. The benefit of expressing $1/\tau^0_{\lambda}$ in this way, is that $W_{\lambda}$ tells us how much scattering is possible, while $\langle |V_{\lambda}|^2 \rangle$ informs us on how likely the scattering is to occur -- the effect of the selection rules are reflected in $\langle |V_{\lambda}|^2 \rangle$. While the unconverged ($1/\tau^0_{\lambda}$) and converged ($1/\tau_{\lambda}$) scattering rates are quite different, analyzing both $\langle |V_{\lambda}|^2 \rangle$ and $W_{\lambda}$ can provide insights into the differences between graphite and graphene. A similar approach was discussed in Ref.~\cite{Singh2011a}.

\begin{figure*}	
	\includegraphics[width=17cm]{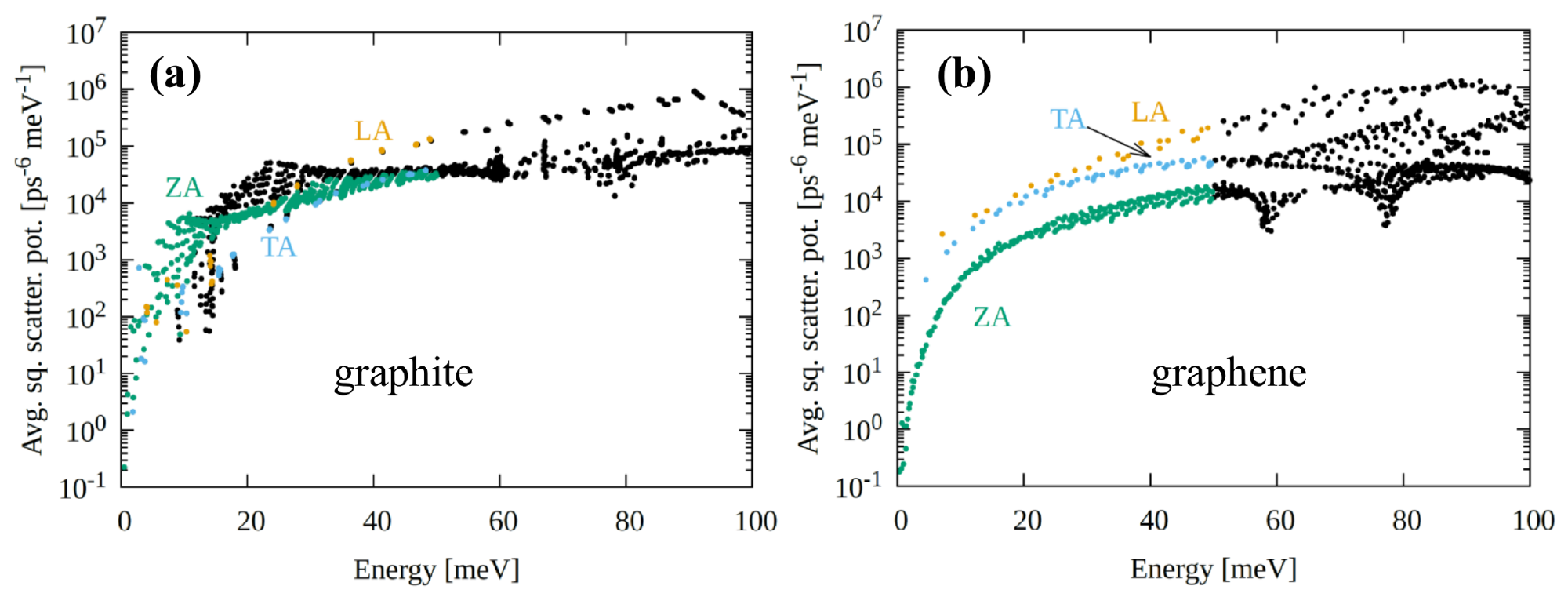}
	\caption{Average squared scattering potential, $\langle |V_{\lambda}|^2 \rangle$, versus phonon energy for graphite (a) and graphene (b). $T$\,=\,300 K.} \label{fig:scatpot}
\end{figure*}

Fig.~\ref{fig:wps} compares the weighted phase space of graphite and graphene (the stretched cases are between both limiting cases). $W_{\lambda}$ decreases with increasing energy, $\hbar \omega_{\lambda}$, which is mostly due to a factor of $1/\omega_{\lambda}$ in the definition (see Eqns.~(\ref{eq:wps1})-(\ref{eq:wps2})). There is, however, a marked increase in $W_{\lambda}$ below $\approx$\,20~meV that occurs because flexural phonons can participate in more scattering processes compared to other phonons. Because the ZA dispersion is relatively flat there is a wide range of possible $\mathbf{q}$, over a small energy range, that can accommodate possible 3-phonon processes.

\begin{figure*}	
	\includegraphics[width=17cm]{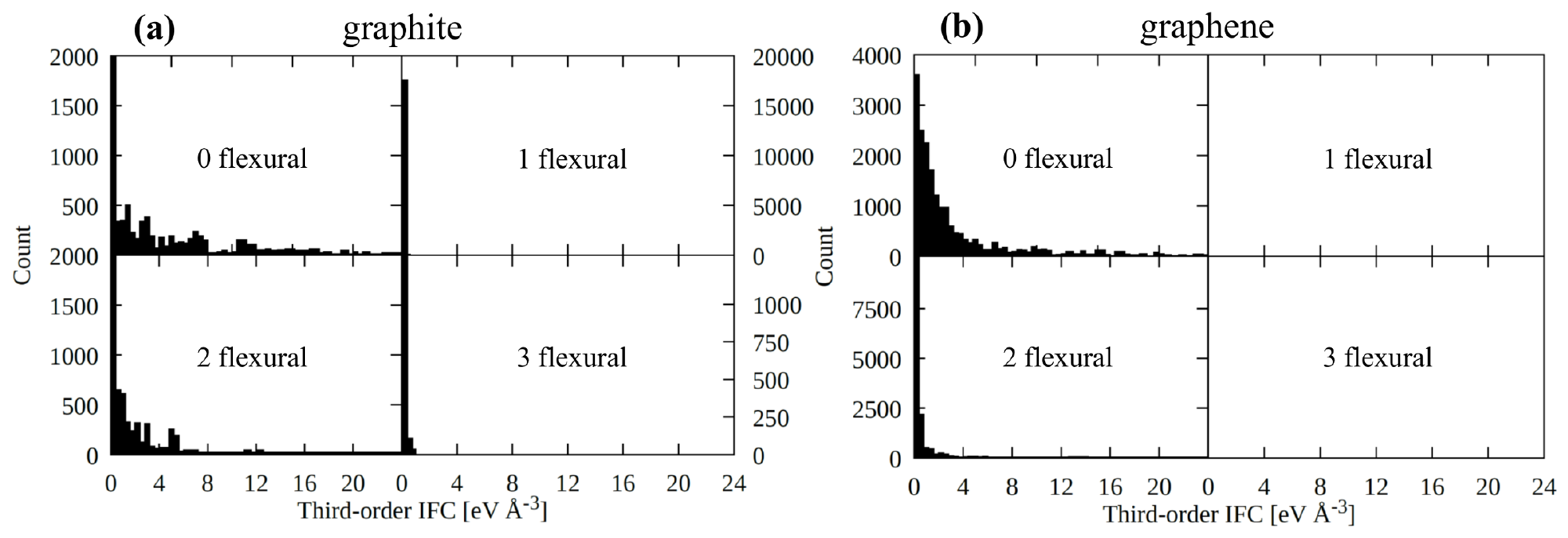}
	\caption{Histogram of third-order IFCs separated into processes involving 0, 1, 2 or 3 flexural phonons, for graphite (a) and graphene (b).} \label{fig:ifcs}
\end{figure*}

We find that $W_{\lambda}$ is slightly larger with graphene, with a sharper rise below 10~meV. This occurs because $W_{\lambda}$ scales with the equilibrium occupation of the scattering phonons, and scales inversely with the energies of the scattering phonons $1/(\omega_{\lambda'}\omega_{\lambda''})$ (see Eqns.~(\ref{eq:wps1})-(\ref{eq:wps2})) -- both these factors are enhanced when the phonons participating in scattering have lower energies, which is the case for the ZA phonons in graphene versus the ZO' in graphite. So far, the weighted phase space suggests that graphene has the capacity for more 3-phonon scattering than graphite.

In Fig.~\ref{fig:scatpot} we show the average squared scattering potential, $\langle |V_{\lambda}|^2 \rangle$, for graphite and graphene. With graphene the scattering strength of the ZA phonons is very small at low energies, as expected from the selection rules. In the case of graphite, we observe more spread in the average scattering potential near 10~meV and 25~meV, resulting from the states with $q_z$\,$\neq$\,0. Interestingly, the ZA modes in graphite also possess some of the lowest $\langle |V_{\lambda}|^2 \rangle$ values, especially below 10~meV, which nearly mimic those of graphene. There are also noticeable dips in the average scattering potential between 10-15~meV, arising from the ZO' modes. Since the selection rules do not strictly apply to graphite, we would have expected much larger $\langle |V_{\lambda}|^2 \rangle$ compared to graphene. Next, we analyze the third-order IFCs to understand the impact of the selection rules.

Fig.~\ref{fig:ifcs} presents a histogram of the third-order IFCs for scattering processes involving 0, 1, 2 or 3 flexural phonons, for graphite and graphene. Graphene shows a distribution of IFC values for processes with 0 or 2 flexural phonons, but the IFCs are exactly zero when 1 or 3 flexural phonons are participating (as expected from the selection rules). With graphite we also observe a range of IFC values for processes containing 0 or 2 flexural phonons, however with 1 or 3 flexural phonons roughly 60\%  of the IFCs are zero with the remaining clustered near zero.

As discussed in Section~\ref{sec:selectrules}, and in Ref.~\cite{Lindsay2011}, the selection rules are not applicable to graphite because a mirror symmetry operation does not map each atom onto itself (unlike with graphene). However, graphite is mirror symmetric, and when all three atoms ($ijk$) determining the third-order IFC $\Phi_{ijk}^{\alpha\beta\gamma}$ are located within the same atomic sheet (denoted here as {\it intra}-layer IFC), then the third-order IFCs vanish for odd number of flexural phonons $\Phi_{\rm intra}^{1z,3z}$\,=\,0, (superscript denotes displacements containing 1 or 3 $z$-components associated with the flexural phonons). When one or more of the three atoms ($ijk$) are located in different layers ({\it inter}-layer IFC), the $\Phi_{\rm inter}^{\alpha\beta\gamma}$ are not required to vanish. In practice, however, the $\Phi_{\rm inter}^{\alpha\beta\gamma}$ are small because they arise from the weak van der Waals coupling. This expains why the IFCs in graphite are similar to those in graphene -- the intra-layer IFCs obey the selection rules and the remaining inter-layer IFCs are non-zero but small (for any number of flexural phonons).

We also note that while the eigen-displacements of graphite are not strictly required to be oriented parallel or perdendicular to $z$, like in the case of graphene, the weak inter-layer coupling only slightly alters the displacement angle; 95\% of the eigen-displacements are within 1$^{\circ}$ of the $z$ axis or the $x$-$y$ plane.

Summarizing our findings for the scattering characteristics; graphene has larger $W_{\lambda}$ than graphite, and shows very small average scattering potential, due to the selection rules. Unexpectedly, graphite displays $\langle |V_{\lambda}|^2 \rangle$ as low as graphene's, with its third-order IFCs approaching zero when an odd number of flexural phonons participate in scattering.

\section{Discussion}
\label{sec:discussion}
Having analyzed graphite, stretched graphite and graphene, next we highlight the key features that influence thermal transport in the crossover from 3D to 2D and identify which are likely to be shared with other van der Waals materials. We find that the thermal conductivity of graphene is 700~W/m-K larger than that of graphite (see Fig.~\ref{fig:cumulk}a). The source of this difference was expected to be the selection rules that reduce scattering in graphene. While the flexural phonons in graphene do benefit from slightly lower scattering rates than graphite, the changes in phonon dispersion plays an important role.

The low-energy ZO' branch in graphite has characteristics that reduce $\kappa$ compared to the ZA branch in graphene from which they originate -- the ZO' modes arise from the inter-layer coupling between ZA modes, as graphene sheets are brought closer to form graphite. For the following discussion it is convenient to refer to the energy-resolved thermal conductivity, written as $\kappa(E)$\,=\,$D(E)\,g(E)\,[E\,\partial f^0/\partial T]$, where $D(E)$ and $g(E)$ are the energy-resolved thermal diffusivity and DOS, respectively (see Eq.~(\ref{eq:kappaE2}) in Appendix~\ref{app:kappa}).

First, the energy of the ZO' phonons are pushed up compared to the ZA phonons. In terms of the (monotonically decreasing) factor $E\,\partial f^0/\partial T$, any increase in phonon energy will result in a lower contribution to $\kappa(E)$. Second, the average velocity of the phonons, at a given energy, is lower due to the spread in velocities from the states with $q_z$\,$\neq$\,0 (see Fig.~\ref{fig:invelo}). This lowers $\kappa(E)$ since the thermal diffusivity $D(E)$ is proportional to the velocity squared (see Eq.~\ref{eq:diff}). Third, the phonon DOS of graphite (or any 3D material) smoothly decreases to zero as $E$\,$\rightarrow$\,0, while the DOS of graphene is a constant at low energy because of the flexural phonons. The constant DOS of graphene gives a significant contribution to $\kappa(E)$ near zero energy, compared to graphite. This has a distinctive effect on $\kappa^{\rm cumul}(E)=\int_0^E \kappa(E')\,dE'$ below 20~meV, where graphene shows a rapid increase versus graphite (see Fig.~\ref{fig:cumulk}a), which accounts for the 700~W/m-K difference in converged thermal conductivities.

The modifications to the phonon dispersion in the crossover from 3D to 2D, and their influence on $\kappa$, are likely to exist in many other van der Waals materials. As the isolated layers are brought closer together the energy of the acoustic phonons will increase, due to the inter-layer coupling, to form low-lying optical phonons. These optical phonons, corresponding to the atomic layers oscillating relative to each other, will have a similar effect on $\kappa$ as discussed above.

While the selection rules are unique to one-atom-thick materials, such as graphene, we find that 3D layered materials may also in general benefit from reduced 3-phonon scattering. As discussed in the previous section, the third-order IFCs for which the three displaced atoms are in the same $x$-$y$ plane vanish for odd number of $z$ displacements, $\Phi_{\rm intra}^{\rm 1z,\,3z}$\,=\,0, if that plane has a mirror symmetry. The third-order IFCs for which the atoms are in different layers, separated by a van der Waals gap, are likely to be small, $\Phi_{\rm inter}$\,$\approx$\,0. These characteristics will reduce scattering and should be present in other van der Waals materials. Note that in the case of graphite all third-order IFCs fall within both aforementioned cases, which will not be true for materials with layers thicker than one atom (e.g. MoS$_2$ or black phosphorous). Thus, we expect the strongest reduction in 3-phonon scattering when the layered materials are comprised of single-atom sheets and have very weak inter-layer coupling.

\section{Conclusions}
\label{sec:conclusions}
The purpose of this study was to understand how the thermal transport properties of van der Waals materials are modified when varying from bulk to monolayer. As a case study we examined graphite and graphene, using DFT in conjunction with the phonon BTE, and simulated the transition from 3D to 2D by continuously pulling apart the atomic layers and thereby controlling the inter-layer coupling.

We calculated the room-temperature $\kappa$ to be 2730~W/m-K for graphene, and 2030~W/m-K (in-plane) and 6.7~W/m-K (cross-plane) for graphite, with the stretched cases monotonically connecting both limits. The cumulative thermal conductivity revealed that most of the difference between graphite and graphene occurs at low-energy, between 0-20~meV. To understand what influences $\kappa$, we analyzed the changes in phonon dispersion and scattering properties.

In terms of phonon dispersion, an important feature is the creation of low-lying optical flexural (ZO') modes in graphite, corresponding to the monolayers oscillating relative to each other. As the inter-layer coupling decreases the ZO' phonons drop in energy and become the acoustic flexural (ZA) modes in graphene. We find that the evolution from ZA in graphene to ZO' in graphite reduces $\kappa$, due to lower phonon occupation, in-plane velocities and DOS. These changes play a significant role in the reduction of $\kappa$ when going from graphene to graphite -- such features are likely shared with other van der Waals materials.

\begin{figure*}	
	\includegraphics[width=13cm]{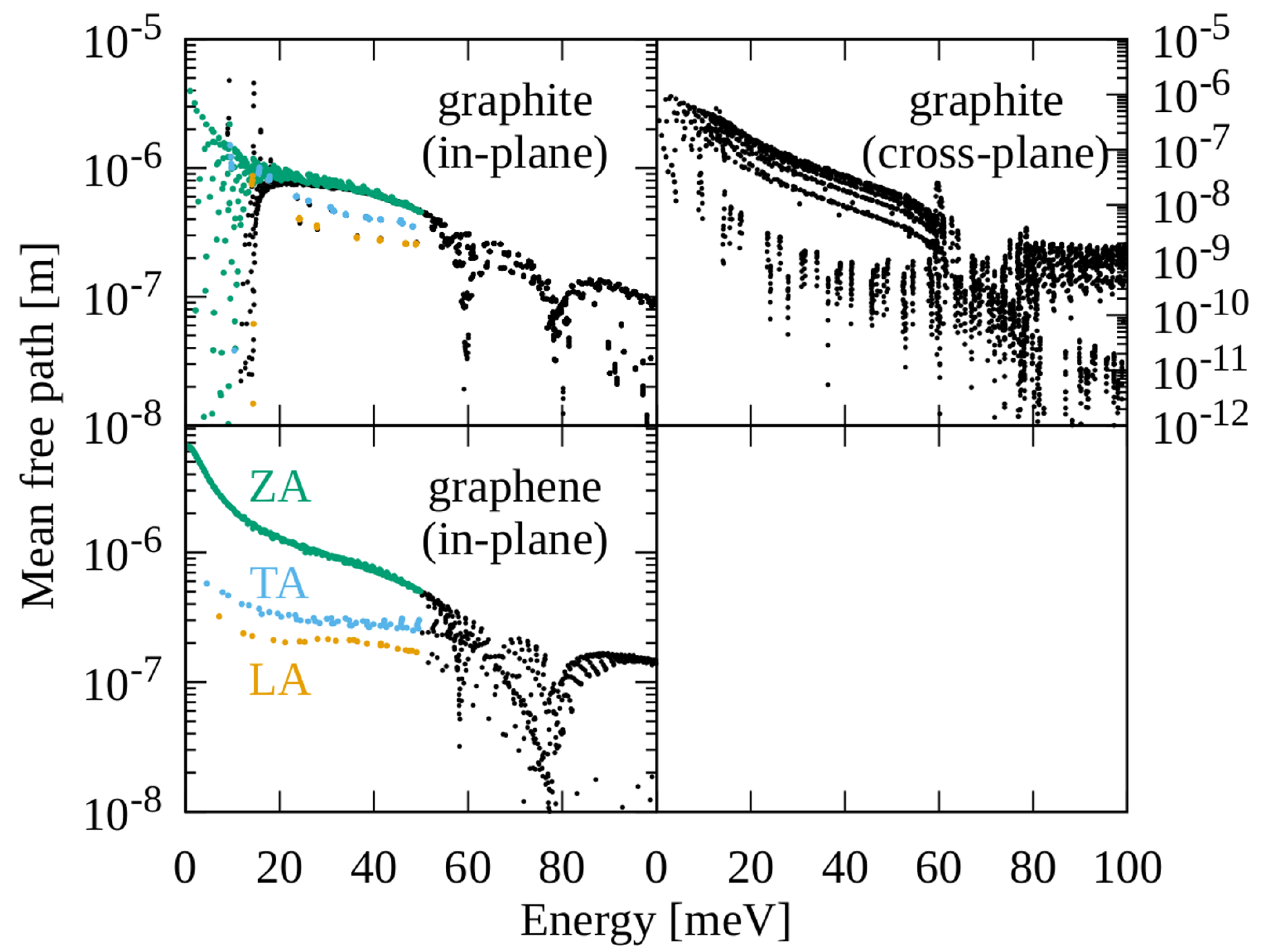}
	\caption{Mean free paths, $|v_{\lambda}^{||,\perp}|\tau_{\lambda}$, versus phonon energy for graphite (in-plane and cross-plane) and graphene, calculated from the converged scattering lifetimes. $T$\,=\,300~K.} \label{fig:mfp}
\end{figure*}

In terms of scattering properties, the 3-phonon scattering rates for graphite and graphene were found to be similar. We expected significantly lower scattering in graphene due to the selection rules. Surprisingly, the third-order IFCs of graphite displayed similar behavior to graphene, with near-zero values for processes involving an odd number of flexural phonons. By separating the IFCs into intra- and inter-layer components, we explained that the intra-layer IFCs adhere to the selection rules and argued that the remaining inter-layer IFCs should be tiny since they depend on weak inter-layer coupling. Due to both these effects that reduce the IFCs, the scattering rates of graphite are comparable to those of graphene. These properties of the third-order IFCs may generally result in lower phonon-phonon scattering in 3D van der Waals materials compared to other non-layered 3D materials.

%

\begin{acknowledgments}
This work was partially supported by DARPA MATRIX (Award No. HR0011-15-2-0037) and NSERC (Discovery Grant RGPIN-2016-04881). P. S. acknowledges support from an NSERC Canada Graduate Scholarship and Nova Scotia Graduate Scholarship.  
\end{acknowledgments}

\appendix
\section{Selection rules and scattering matrix elements}
\label{app:selectrules}
We evaluate the 3-phonon scattering matrix elements to demonstrate the previously reported selection rules in graphene \cite{Lindsay2010,Lindsay2011}. In Section~\ref{sec:selectrules}, we showed that the second- and third-order IFCs, $\Phi_{ij}^{\alpha\beta}$ and $\Phi_{ijk}^{\alpha\beta\gamma}$, are zero for processes involving an odd number of $z$-component displacements, where $(ijk)$ denote specific atoms in the material and $(\alpha\beta\gamma)$ each corresponds to a Cartesian displacement direction $x$, $y$ or $z$.

Starting with the aforementioned symmetry property on $\Phi_{ij}^{\alpha\beta}$, one can show that the dynamical matrix decouples the $z$ components from the $x$-$y$ components. For any pair of atoms $(ij)$, the dynamical matrix takes the form:
\begin{align}
	D_{ij}^{\alpha\beta} = \left( \begin{array}{ccc} D_{ij}^{xx} & D_{ij}^{xy} & 0 \\ D_{ij}^{yx} & D_{ij}^{yy} & 0 \\ 0 & 0 & D_{ij}^{zz} \end{array} \right), \label{eq:Dmat}
\end{align}   
where $D_{ij}^{\alpha\beta} \propto \Phi_{ij}^{\alpha\beta}$ \cite{Ashcroft1976}. The full dynamical matrix, containing all atomic pairs, can be reorganized into a similar form in which the elements that depend on $z$ and $x$-$y$ are grouped into two independent sub-blocks. As a result, the eigenvectors (or eigen-displacements) of the dynamical matrix are either parallel or perpendicular to $z$ -- the former being the flexural phonons. This shows that only flexural phonons can have $z$ displacements, which is key to deriving the following property.

The 3-phonon scattering matrix elements are written as \cite{Li2014}:
\begin{align}
   V^{\pm}_{\lambda\lambda'\lambda''} = \sum_{i\in{\rm u.c.}} \sum_{jk} \sum_{\alpha\beta\gamma} \Phi_{ijk}^{\alpha\beta\gamma} \frac{\epsilon_i^{\alpha}(\lambda) \epsilon_j^{\beta}(\lambda') \epsilon_k^{\gamma}(\lambda'')}{\sqrt{M_iM_jM_k}}, \label{eq:Vscat}
\end{align}
where $\pm$ denotes absorption ($+$) and emission ($-$) processes, atom $i$ is located in the unit cell at the origin, the $\mathbf{\epsilon}$'s are the normalized eigen-vectors of the dynamical matrix, and $M_i$, $M_j$ and $M_k$ are the masses of atoms $i$, $j$ and $k$, respectively. Note that conservation of crystal momentum within a reciprocal lattice vector $\mathbf{Q}$ is assumed in Eq.~(\ref{eq:Vscat}): $\mathbf{q}\pm\mathbf{q}'-\mathbf{q}''+\mathbf{Q}=0$. Next, we demonstrate how $V^{\pm}_{\lambda\lambda'\lambda''}$, and hence the scattering rates, vanishes for processes involving an odd number of flexural phonons.

Consider the case of one incident flexural phonon:
\begin{align}
   V^{\pm}_{{\rm flex}\,\lambda'\lambda''} &= \sum_{i\in{\rm u.c.}} \sum_{jk} \sum_{\beta\gamma\neq z} \Phi_{ijk}^{z\beta\gamma} \frac{\epsilon_i^{z}({\rm flex}) \epsilon_j^{\beta}(\lambda') \epsilon_k^{\gamma}(\lambda'')}{\sqrt{M_iM_jM_k}} \nonumber \\
   &=0, \label{eq:V1flex}
\end{align}
where we used the fact that all flexural phonon displacements are along $z$, and that the third-order IFCs are zero with a single $z$ displacements. Next, let's assume there are three flexural phonons:
\begin{align}
   V^{\pm}_{{\rm flex}\,\,{\rm flex}'{\rm flex}''} &= \sum_{i\in{\rm u.c.}} \sum_{jk} \Phi_{ijk}^{zzz} \frac{\epsilon_i^{z}({\rm flex}) \epsilon_j^{z}({\rm flex}') \epsilon_k^{z}({\rm flex}'')}{\sqrt{M_iM_jM_k}} \nonumber \\
   &=0, \label{eq:V3flex}
\end{align}
where the third-order IFCs are zero with three $z$ displacements. Since the scattering rates are calculated from the scattering matrix elements, this demonstrates that the scattering rates for processes involving an odd number of flexural phonons are zero.

\section{Weighted phase space and average scattering potential}
\label{app:wps_avpot}
Within the relaxation time approximation, the 3-phonon scattering rates $1/\tau^0_{\lambda}$ are expressed as \cite{Li2014}:
\begin{align}
    \frac{1}{\tau^{0,\pm}_{\lambda}} = &\frac{h}{16N_q} \sum_{\lambda' n''} \left\{ \begin{array}{c} 2(f^0_{\lambda'}-f^0_{\lambda''}) \\ f^0_{\lambda'}+f^0_{\lambda''}+1 \end{array} \right\} |V^{\pm}_{\lambda\lambda'\lambda''}|^2 \nonumber \\ & \times \frac{\delta(\omega_{\lambda}\pm\omega_{\lambda'}-\omega_{\lambda''})}{\omega_{\lambda}\omega_{\lambda'}\omega_{\lambda''}}, \label{eq:scat_rta}
\end{align}
where `$+$' and `$-$' correspond to absorption and emission processes (along with the upper and lower components in curly brackets), which are additive $1/\tau^0_{\lambda}=1/\tau^{0,+}_{\lambda}+1/\tau^{0,-}_{\lambda}$. ($1/\tau^0_{\lambda}$ corresponds to the zeroth-order approximation in the self-consistent calculation of the phonon BTE.) Conservation of crystal momentum within a reciprocal lattice vector $\mathbf{Q}$ determines the wavevector $\mathbf{q}''=\mathbf{q}\pm\mathbf{q}'+\mathbf{Q}$. $V_{\lambda\lambda'\lambda''}$ are the 3-phonon scattering matrix elements computed from the third-order IFCs (see Eq.~(\ref{eq:Vscat})).

The weighted phase space, as calculated in the ShengBTE code, is defined as \cite{Li2015}:
\begin{align}
    W^{\pm}_{\lambda} &= \frac{1}{N_q} \sum_{\lambda' n''} W^{\pm}_{\lambda\lambda'\lambda''}, \label{eq:wps1} \\
    W^{\pm}_{\lambda\lambda'\lambda''} &= \frac{1}{2} \left\{ \begin{array}{c} 2(f^0_{\lambda'}-f^0_{\lambda''}) \\ f^0_{\lambda'}+f^0_{\lambda''}+1 \end{array} \right\} \frac{\delta(\omega_{\lambda}\pm\omega_{\lambda'}-\omega_{\lambda''})}{\omega_{\lambda}\omega_{\lambda'}\omega_{\lambda''}}, \label{eq:wps2}
\end{align}
where the absorption and emission processes are additive $W_{\lambda}=W^+_{\lambda}+W^-_{\lambda}$. Eqns.~(\ref{eq:wps1})-(\ref{eq:wps2}) are equivalent (within a constant factor) to Eq.~(\ref{eq:scat_rta}) with the 3-phonon scattering matrix elements removed, and thus sums all possible processes regardless of $V_{\lambda\lambda'\lambda''}$. Using the above expressions for $1/\tau^0_{\lambda}$ and $W_{\lambda}$, we can define the following:
\begin{align}
    \frac{1}{\tau^0_{\lambda}} &= \frac{h}{8} \langle |V_{\lambda}|^2 \rangle W_{\lambda}, \label{eq:scat_VW2} \\
   \langle |V_{\lambda}|^2 \rangle &\equiv \frac{\sum_{\lambda'\lambda''} ( W^+_{\lambda\lambda'\lambda''}|V^+_{\lambda\lambda'\lambda''}|^2+W^-_{\lambda\lambda'\lambda''}|V^-_{\lambda\lambda'\lambda''}|^2 )}{\sum_{\lambda'\lambda''} ( W^+_{\lambda\lambda'\lambda''}+W^-_{\lambda\lambda'\lambda''})}, \label{eq:avscatpot}
\end{align} 
where $\langle |V_{\lambda}|^2 \rangle$ is the average squared scattering potential. The ShengBTE code outputs both $1/\tau^0_{\lambda}$ and $W_{\lambda}$, from which we calculate $\langle |V_{\lambda}|^2 \rangle$ using Eq.~(\ref{eq:scat_VW2}). From the definitions, the units are $1/\tau^0_{\lambda}$\,=\,[1/s], $W_{\lambda}$\,=\,[s$^4$],  and $\langle |V_{\lambda}|^2 \rangle$\,=\,[J$^{-1}$s$^{-6}$].

\section{Energy-resolved thermal conductivity}
\label{app:kappa}
The thermal conductivity given by Eq.~(\ref{eq:kappa}) can be rewritten in terms of the energy-resolved thermal conductivity, $\kappa(E)$:
\begin{align}
	\kappa^{\alpha\beta} &= \int_0^{\infty} \kappa^{\alpha\beta}(E)\,dE, \label{eq:kappa2} \\
	\kappa^{\alpha\beta}(E) &=  \sum_{\lambda} v_{\lambda}^{\alpha} v_{\lambda}^{\beta} \tau_{\lambda} E_{\lambda}\frac{\partial f^0_{\lambda}}{\partial T} \,\frac{\delta(E-E_{\lambda})}{\Omega N_q}. \label{eq:kappaE} 
\end{align}
Eq.~(\ref{eq:kappaE}) can be expressed alternatively by definining the following quantities:
\begin{align}
	D^{\alpha\beta}(E) &= \langle v^{\alpha}v^{\beta}\tau \rangle  \nonumber \\
	&=  \frac{\sum_{\lambda} v_{\lambda}^{\alpha} v_{\lambda}^{\beta} \tau_{\lambda} \,\delta(E-E_{\lambda})}{\sum_{\lambda} \delta(E-E_{\lambda})}, \label{eq:diff} \\
	g(E) &= \frac{1}{\Omega N_q} \sum_{\lambda} \,\delta(E-E_{\lambda}), \label{eq:dos} \\
	C_V(E) &= g(E)\left[ E\,\frac{\partial f^0}{\partial T}\right], \label{eq:CVE}
\end{align}
which are the energy-resolved thermal diffusivity, DOS and heat capacity. Using Eqns.~(\ref{eq:diff})-(\ref{eq:CVE}) in Eq.~(\ref{eq:kappaE}) we obtain:
\begin{align}
	\kappa^{\alpha\beta}(E) &= D^{\alpha\beta}(E)\,C_V(E) \nonumber \\
	&= D^{\alpha\beta}(E)\,g(E)\left[ E\,\frac{\partial f^0}{\partial T}\right], \label{eq:kappaE2} 
\end{align}
which shows that the energy-resolved thermal conductivity is proportional to the energy-resolved diffusivity, DOS and the material-independent function $[E\,\partial f^0/\partial T]$.


\end{document}